\documentclass[twocolumn]{emulateapj}
\usepackage{amsmath}
\usepackage{graphicx}
\usepackage{multirow}
\usepackage{color}
\usepackage{latexsym}
\usepackage{amssymb}
\usepackage{epsfig}

\begin{document}
\title{Superluminous light curves from supernovae exploding in a dense wind}

\author{Sivan Ginzburg and Shmuel Balberg}

\affil{Racah Institute of Physics, The Hebrew University, Jerusalem 91904, Israel}

\begin{abstract}

Observations from the last decade have indicated the existence of a general class of superluminous supernovae (SLSNe), in which the peak luminosity exceeds $10^{44}$ erg s$^{-1}$. Here we focus on a subclass of these events, where the light curve is also tens of days wide, so the total radiated energy is order $10^{51}$ erg. If the origin of these SLSNe is a core-collapse-driven explosion of a massive star, then the mechanism which converts the explosion energy into radiation must be very efficient (much more than in typical core collapse SNe, where this efficiency is of order one percent). We examine the scenario where the radiated luminosity is due to efficient conversion of kinetic energy of the ejected stellar envelope into radiation by interaction with an optically thick, pre-existing circumstellar material (CSM), presumably the product of a steady wind from the progenitor. We base the analysis on a simple, numerically solved, hydrodynamic diffusion model, which allows us to identify the qualitative behavior of the observable light curves, and to relate them to the parameters of the wind. We specifically show that a wide and superluminous supernova requires the mass of the relevant wind material to be comparable to that of ejected material from the exploding progenitor. We find the wind parameters which explain the peak luminosity and width of the bolometric light curves of three particular SLSNe, namely, SN 2005ap, SN 2006gy, and SN 2010gx, and show that they are best fitted with a wind that extends to a radius of order $10^{15}$ cm. These results serve as an additional indication that at least some SLSNe are powered by interaction of the ejected material with a steady wind of similar mass.
\end{abstract}


\section{Introduction}
\label{sec:Introduction}

The discovery of extremely luminous transients in the last years led to their classification as superluminous supernovae, or SLSNe \citep{Quimby2007,Quimby2011,Pastorello2010,Smith2010,GalYam2012}. In many of these events, the light curve is tens of days wide, so the total radiated energy is of order $\sim 10^{51}$ erg. The radius of the photosphere at peak luminosity is $R_{\rm ph}\gtrsim 10^{15}$ cm, as inferred from the observed temperature, assuming blackbody emission. These high values of radiated energy challenge our understanding of the energy source and the conversion mechanism of available energy to radiated energy.

The standard model of core collapse supernovae is that the explosion initiates a 
shock wave which propagates through the progenitor and deposits about half of the explosion energy as thermal energy and half as kinetic energy of the ejecta (for a strong shock in an ideal gas). Red supergiant (RSG) supernovae progenitors have a typical initial radius of $R_*\sim 10^{13}-10^{14}$ cm. For such a progenitor, as the shock reaches the stellar surface the star is highly opaque, so most of the thermal energy is transformed to kinetic energy of the ejecta via adiabatic expansion, and the fraction of thermal energy which escapes as radiation when the expanding ejecta becomes transparent is roughly $R_*/R_{\rm ph}$ \citep{Arnett1996}. Other progenitors are even more compact. Therefore, in order to explain the radiated energy in SLSNe, we must either assume an energy source much larger than $\sim 10^{51}$ erg (typical for core collapse SNe), or find a more efficient mechanism of transforming the explosion energy to radiation.

There have been several suggestions of efficient mechanisms for converting the explosion energy into emerging radiation \citep{SmithMcCray2007,Woosley2007,ChevalierIrwin2011,Chatzopoulos2012}. These suggestions are all based upon extensive pre-explosion mass-loss - a wind - by the progenitor star which expels matter to large radii, comparable with $R_{\rm ph}$. In principle this can be the result of a burst of mass loss or of a steady wind. 
This matter contributes to the radiated energy in two complimentary manners. The extended mass defines an effective radius $R_*^{\rm eff}>R_*$, over which the shock deposits the explosion energy (essentially, creating a "bloated" star). With this larger radius, less of the thermal energy is lost through adiabatic expansion, and a larger fraction $R_*^{\rm eff}/R_{\rm ph}$ of this energy emerges as radiation. In addition, this material interacts with the expanding ejecta. This interaction converts kinetic energy of the expanding ejecta back to thermal energy via shocks propagating forward through the wind material and backward through the ejecta \citep{GalYam2012}. In the roughest approximation, this interaction between the ejecta and the wind can be regarded as a plastic collision between an ejecta mass $M_{\rm ej}$ moving at a certain velocity and a "stationary" wind mass $M_w$. In this approximation, a fraction $\sim M_w/\left(M_{\rm ej}+M_w\right)$ of the ejecta kinetic energy will be converted to shock energy. If these shocks deposit the energy at a radius $\sim R_{\rm ph}$ then it will escape as radiation without suffering significant adiabatic losses. 
We conclude that an excess mass of $M_w\sim M_{\rm ej}$ extending to a radius $\sim R_{\rm ph}$ can convert a substantial portion of the explosion energy to radiated energy and therefore explain the high luminosities of SLSNe with a conventional $\sim 10^{51}$ erg energy source.   

Most of the previous works on supernova explosions in a dense mass loss \citep{SmithMcCray2007,Smith2010,Balberg2011,ChevalierIrwin2011,Chatzopoulos2012} used simple approximations which allow analytical order of magnitude estimations. Some of these works also drew conclusion from the self-similar solutions of \citet{Chevalier1982}, which describe the interaction of an expanding ejecta with a wind in the limit where the wind mass is small, $M_{\rm ej}\gg M_w$, so only the outer layers of the ejecta interact with the wind, and $M_{\rm ej}$ does not create a natural scale in the dynamics. For a more accurate relation between parameters of the star-wind system and the observed light curve, a self-consistent hydrodynamic calculation is required. Early numerical calculations of supernova explosion into a dense wind have been carried out by \citet{FalkArnett1973,FalkArnett1977}, and more systematically by \citet{Moriya2011}. These works do not cover well the regime $M_w\sim M_{\rm ej}$ which is of interest in the current work, in order to achieve a high efficiency in terms of generating a luminous light curve. One exception is the very recent work of \citet{Moriya2012new}, who conducted several numerical calculations of a light curve for the specific case of SN 2006gy: they considered the relevant mass regime for a similar but different case of a shell set at a standoff distance $R_{\rm ph}$ from an exploding star.

In principle the circumstellar material (CSM) can be the result of a steady wind or a burst of mass loss during the last stages of the stars evolution. When the wind is steady the final configuration is a star enclosed in a continuous envelope, while a burst ends with a thin shell of matter situated at some distance from the star. In this work we focus on the steady wind scenario, for which the density profile is well defined ($r^{-2}$ dependence), and can be motivated by various aspects of late stages of stellar evolution \citep[see][]{QS2012}. We use hydrodynamic diffusion calculations to conduct a general survey of the relations between the light curve and the progenitor and wind parameters. We focus on the particular case of a steady wind generated during the last stages of the progenitor star's evolution, so the wind essentially extends from the surface of the star. Our goal is to keep the model as simple as possible, and therefore we adopt many simplifying approximations, which while still sufficient to investigate the key features of light curves in the interacting ejecta and wind scenario, are also simple enough to be instructive and the results are easily understood. The simplifications we applied are detailed below.

The outline of the paper is as follows. In \S~\ref{sec:Progenitors} we present our model for general SLSN progenitor systems. In \S~\ref{sec:QualitativePicture} we consider the qualitative relations between the system parameters and the observed light curve, and analytically examine various limits of the problem. The hydro-diffusion code is described in \S~\ref{sec:the_code} and numerical results are presented in \S~\ref{sec:numerical_results}. The observed light curves of three specific candidates for an interacting ejecta and wind scenario, SN 2005ap, SN 2006gy, and SN 2010gx are analyzed and reproduced in \S~\ref{sec:observations}. We summarize our conclusions in \S~\ref{sec:conclusions}.

\section{Progenitor Systems}
\label{sec:Progenitors}

In our scenario the progenitor systems is composed of two parts: the gravitationally bound compact star, and the outer material which is created by pre-explosion mass loss. Both are characterized by their total masses and respective density profiles.
Since the supernova explosion induces temperatures much higher than the initial star temperatures, and ejecta velocities much higher than the initial wind velocity, both the star and wind are approximated as initially stationary and cold. The explosion can then be treated as an instantaneous release of thermal energy $E$ at the center of the star ($r=0$).

For systems in which the photosphere lies at $R_{\rm ph}\gg R_*$ and $M_{\rm ej}\sim M_w$ the details of the structure of the progenitor star are unimportant in terms of the resulting light curve, since the star can be considered as a point object. None the less, we do want to allow also for calculations of systems where the wind mass is much smaller than the ejected mass, mainly for comparison with other works which investigated these cases \citep[see e.g.][]{Moriya2011}. We follow \citet{Matzner1999} by modeling the progenitor star by a polytrope with radius $R_*$, mass $M_*$, and polytropic index $n$. $n=3/2$ is suitable for convective envelopes, like those in RSGs, while $n=3$ is used to describe radiative envelopes, such as those in blue supergiants (BSGs). We approximate the star by applying the Lane-Emden equation to the whole star, not only to the outer envelope. Thus, $R_*$,$M_*$, and $n$ completely determine the initial density profile of the star. Note that since the Lane-Emden equation is applicable only for the envelope, $M_*$ actually denotes the envelope mass, and not the star mass. All these approximations are valid since the inner structure of the mantle has little effect on the shock wave propagation. The mantle can be considered as a point mass which is added artificially \citep[see][]{Matzner1999}. Thus, in terms of size, a $15M_\sun$ progenitor should actually be modeled by a $M_*\sim 10M_\sun$ structure, to allow for the $\sim 5M_\sun$ mantle.

For the wind density profile, we follow \citet{Balberg2011} and \citet{ChevalierIrwin2011} and consider a steady wind with mass loss rate $\dot{M}$ and wind velocity $v_w$. The resulting density profile is
\begin{equation}\label{eq:wind_density}
\rho(r)=\frac{\dot{M}}{4\pi r^2v_w}\equiv Kr^{-2}.
\end{equation}
We impose this wind density profile by continuously attaching it to the polytropic structure of the star at a radius where the polytrope and equation \eqref{eq:wind_density} coincide. For the parameters chosen in this work, this wind base radius is essentially at $R_*$. 

The wind is assumed to continue to an outer radius $R_w$ so that its total mass is
\begin{equation}\label{eq:wind_mass}
M_w=\int^{R_w}_{R_*}{4\pi r^2Kr^{-2}dr}=4\pi K\left(R_w-R_*\right).
\end{equation}
For $R_w\gg R_*$, which is the case of interest in this work, equation \eqref{eq:wind_mass} is reduced to
\begin{equation}\label{eq:wind_mass1}
M_w\approx 4\pi KR_w.
\end{equation} 
For numerical convenience, we smooth the wind edge at $R_w$. Instead of cutting the wind abruptly at $R_w$, we extend the wind to a larger radius $r=\alpha R_w$ with $\alpha>1$ (we usually choose $\alpha\sim 1.2$) and change the wind density profile from equation \eqref{eq:wind_density} to
\begin{equation}
\rho(r)=Kr^{-2}e^{{-\left(r/R_w\right)}^k}
\end{equation}
with $k\sim 10$ providing a sharp power law. The details of this smoothing have a minor effect on the results. 

Since the light curve is mostly dominated by the structure of the wind material, we mostly survey the wind parameters and fix the parameters of the star. We choose a star with $R_*=10^{13}$cm, $M_*=15M_\sun$, and $n=3/2$. These values are representative of a typical RSG. We study the case $R_w\gg R_*$, for which the value of $R_*$ has little effect on the light curve.
The value of $n$ (3/2 or 3) also does not effect the results in the regime of interest. However, the value of $M_*$ does has a significant effect on the light curve, and the choice of  $15M_\sun$ is arbitrary. Some consequences of changing $M_*$ are discussed in \S~\ref{sec:observations}.

The motivations described in \S~\ref{sec:Introduction} lead to the choice $R_w\sim R_{\rm ph}\sim 10^{15}$ cm (the relation $R_w\sim R_{\rm ph}$ is justified in \S~\ref{sec:QualitativePicture}) and $M_w\sim M_{\rm ej}\sim M_*$, which, using equation \eqref{eq:wind_mass1}, sets $K\sim 10^{18} \textrm{ g cm}^{-1}$.

To summarize, our progenitor system is thus modeled by 4 parameters: the star mass $M_*$, the explosion energy $E$, the wind outer radius $R_w$, and the wind density coefficient $K$ from equation \eqref{eq:wind_density}. After fixing the star mass, we are left with 3 parameters which determine the properties of the light curve. 

\section{Qualitative Picture and Extreme Limits}
\label{sec:QualitativePicture}

The supernova explosion initiates by a strong shock wave, which when close to the edge of the star becomes radiation dominated.  
Diffusion of energy carried by radiation causes the shock to develop a finite width with optical depth $\delta\tau\sim\beta_{\rm sh}^{-1}\equiv c/v_{\rm sh}$ \citep{Weaver1976}, where $v_{\rm sh}$ is the shock velocity and $c$ the speed of light. This optical depth can be intuitively understood by equating the hydrodynamical time scale and the diffusion time scale over the shock front. At large optical depths from the surface, $\tau\gg\beta_{\rm sh}^{-1}$, the shock wave can still be treated as an ideal discontinuity, and diffusion can be neglected. When the shock approaches the surface, and $\tau\sim\beta_{\rm sh}^{-1}$, energy can escape by diffusion to the surface, and and the shock dissolves; the purely hydrodynamical (with diffusion neglected) solutions are no longer valid. At this stage the shock is said to "breakout".

Shock breakout through the outer layers of a bare star has been discussed in several works. These rely on the self similar solution found by \citet{Sakurai1960} for a planar shock propagating through the steeply declining density profile at the edge of the star; see \citet{Matzner1999} for a complete derivation. \citet{Sapir2011} expanded this self similar solution to include diffusion and thus calculate a light curve under these conditions, again - for the plane-parallel case. In the presence of an extended wind, due to a pre-explosion mass loss from the progenitor, shock breakout must be considered in spherical symmetry. As the outgoing ejecta plows through the wind and slows down, it drives a forward shock through the wind and a reverse shock that propagates back into the ejecta, and it is the combined shock profile which eventually breaks out as it reaches a low optical depth region in the wind. Our focus is on the case where $M_w\sim M_{\rm ej}$, so the entire ejecta participates in the shock breakout, and the light curve is the result of the breakout through the opaque wind \citep{ChevalierIrwin2011}.  

While no self-similar solution exists for the general case, we can still quantify several important conclusions about the typical time scale of the light curve. We can also draw some insight from the two extreme limits where $M_w\gg M_{\rm ej}$ and $M_w\ll M_{\rm ej}$. If the wind mass is much larger than the ejecta mass, the dynamics are similar to the Sedov-Taylor explosion, but with a power law ambient density instead of a constant one. In the opposite extreme limit ($M_w\ll M_{\rm ej}$), the self-similar solution of \citet{Chevalier1982} can be used to model the combined forward and reverse shock. For such a configuration only the initial light curve will be the result of shock breakout through the wind \citep{Balberg2011}, followed by the main light curve driven by the internal energy held by the bulk of the ejecta. In the following we examine the estimates that can be made about the resulting light curve from these solutions and other considerations; the qualitative picture we draw serves to clarify the numerical results presented in the later sections. 

\subsection{Breakout Radius}
\label{subsec:Breakout Radius}

As the forward shock propagates through the wind, it breaks out when it reaches optical depth $\tau\sim c/v_{\rm sh}$. We adopt the assumption of constant opacity,
which is appropriate for electron (Thompson) scattering \citep[a similar assumption was adopted by][]{Arnett1996,ChevalierIrwin2011,Balberg2011,Moriya2011}. The value of $\kappa$ is, of course, composition dependent, ranging between $\kappa\approx 0.2\textrm{ cm}^2\textrm{g}^{-1}$ for Hydrogen free matter and $\kappa\approx 0.4\textrm{ cm}^2\textrm{g}^{-1}$ for pure Hydrogen (we usually used  $\kappa\approx 0.34\textrm{ cm}^2\textrm{g}^{-1}$ which is appropriate for a 70\% Hydrogen composition). In this case of constant opacity the optical depth from a radius $r$ to the edge of the wind is given by
\begin{equation}\label{eq:optical_depth}
\tau(r)=\int_r^{R_w}{\kappa\rho dr}=\int_r^{R_w}{\kappa Kr^{-2}dr}=\kappa K\left(\frac{1}{r}-\frac{1}{R_w}\right).
\end{equation}
Therefore, the shock breaks out at radius $R_{\rm sh}$ which satisfies
\begin{equation}\label{eq:breakout_radius}
\frac{c}{v_{\rm sh}}\sim \kappa K\left(\frac{1}{R_{\rm sh}}-\frac{1}{R_w}\right).
\end{equation}
Following \citet{ChevalierIrwin2011} we denote
\begin{equation}\label{eq:rd}
R_d\equiv\frac{\kappa K v_{\rm sh}}{c},
\end{equation}
and rewrite equation \eqref{eq:breakout_radius} as
\begin{equation}\label{eq:breakout_rd}
\frac{1}{R_{\rm sh}}\approx\frac{1}{R_w}+\frac{1}{R_d}.
\end{equation}

\citet{ChevalierIrwin2011} discussed the breakout at the two limits: 
\begin{equation}\label{eq:rshock_limits}
R_{\rm sh}\approx
\begin{cases}
R_w &R_w\ll R_d \\
R_d & R_w\gg R_d
\end{cases}.
\end{equation}
Our focus is on the intermediate case of $R_d\approx R_w$, which appears to be motivated by observations. Note that as long as the wind mass is of order the ejecta mass, the shock velocity naturally tends to $v_{\rm sh}\sim\sqrt{E/M_*}\sim 3\times 10^8\textrm{ cm s}^{-1}$, so for the parameters chosen in \S~\ref{sec:Progenitors}, $R_d\sim 10^{15}$ cm.

\subsection{Luminosity and Timescale}\label{subsec:Ltdiff}

The main features of the observed light curve are total radiated energy, the peak luminosity and the typical width. The three are connected, of course, through the emission of the thermal energy at shock breakout by photon diffusion through the wind.  Being an integral quantity, the total emitted energy, $E_{\textrm{rad}}$ is expected to follow the plastic collision picture described in \S~\ref{sec:Introduction} giving the simple relation
\begin{equation}\label{eq:e_rad}
E_{\textrm{rad}}\propto E\frac{M_w}{M_{\rm ej}+M_w}\sim E\frac{M_w}{M_*+M_w}.
\end{equation}  

The typical timescale in the light curve must depend not only on integral quantities, but on the details of the wind profile as well. We estimate the time it takes the shock energy to diffuse to the surface following breakout as follows. First, we note that the photons do not diffuse all the way from $R_{\rm sh}$ to the surface, but rather to the radius of the photosphere $R_{\rm ph}$ which is located at $\tau\sim 1$ (specifically $\tau=2/3$ for Eddington's approximation). Using equation \eqref{eq:optical_depth}, $R_{\rm ph}$ is given by
\begin{equation}\label{eq:r_ph}
\frac{1}{R_{\rm ph}}\approx\frac{1}{R_w}+\frac{1}{\kappa K}.
\end{equation}
For the case $R_w\ll R_d<\kappa K$, equation \eqref{eq:r_ph} yields $R_{\rm ph}\approx R_w$. According to equation \eqref{eq:breakout_rd} $R_{\rm sh}\approx R_w$ in this case as well, so the shock breaks out very close to the edge of the wind and we can assume a constant diffusion coefficient
\begin{equation}\label{eq:d_diff_rw}
D\sim\frac{c}{\kappa\rho}\sim\frac{cR_w^2}{\kappa K}
\end{equation}
and a diffusion distance (see equations \eqref{eq:breakout_rd}, \eqref{eq:r_ph}, and assuming $v_{\rm sh}\ll c$) of
\begin{equation}\label{eq:delta_r}
\Delta R=R_{\rm ph}-R_{\rm sh}\approx R_w-R_{\rm sh}=\frac{R_w^2}{R_w+R_d}\approx\frac{R_w^2}{R_d}
\end{equation} 
which results in a diffusion time
\begin{equation}\label{eq:tdiff}
t_d\approx\frac{\Delta R^2}{D}=\frac{R_w^2}{\left(\frac{\kappa K}{c}\right)v_{\rm sh}^2}.
\end{equation} 
In the opposite limit, $R_w\gg R_d$, and so $R_{\rm sh}\approx R_d\ll R_w$. Since the shock velocity is far from relativistic, we also have $R_{\rm sh}\approx\kappa Kv_{\rm sh}/c\ll\kappa K$. These relations imply, according to equation \eqref{eq:r_ph}, that $R_{\rm sh}\ll R_{\rm ph}$. In this case shock breakout evolves through the diffusion of radiation to the photosphere, and the diffusion time can be estimated by considering the change of the density-dependent diffusion coefficient
\begin{equation}\label{eq:tdiff_integral}
t_d\approx\int_{R_{\rm sh}}^{R_{\rm ph}}{\frac{d(r-R_{\rm sh})^2}{D(r)}}\sim
\int_{R_{\rm sh}}^{R_{\rm ph}}{\frac{(r-R_{\rm sh})\kappa\rho dr}{c}}.
\end{equation}
After substituting a $\rho=Kr^{-2}$ density profile the integration yields
\begin{equation}\label{eq:tdiff_integrated}
t_d~\approx\frac{\kappa K}{c}\left(\ln{\frac{R_{\rm ph}}{R_{\rm sh}}}+\frac{R_{\rm sh}}{R_{\rm ph}}-1\right).
\end{equation}
Combining this result with equations \eqref{eq:breakout_rd} and \eqref{eq:r_ph} gives a general expression for the diffusion time for any $R_w$.
By substituting $R_{\rm sh}\approx R_d$ from equation \eqref{eq:rd} and $R_{\rm ph}$ from equation \eqref{eq:r_ph}, and using $R_{\rm sh}\ll R_{\rm ph}$ we summarize the diffusion time at both limits:
\begin{equation}\label{eq:tdiff_both}
t_d\approx
\begin{cases}
\frac{R_w^2}{\left(\frac{\kappa K}{c}\right)v_{\rm sh}^2}  & R_w\ll R_d \\[1.5ex]
\frac{\kappa K}{c}\left[\ln\left(\frac{c}{v_{\rm sh}}\frac{1}{1+\left[\kappa K/R_w\right]}\right)-1\right] & R_w\gg R_d
\end{cases}.
\end{equation}
Equation \eqref{eq:tdiff_both} gives a diffusion time which is monotonically increasing with $R_w$ in both limits, but more mildly at large $R_w$. For $R_w\gg\kappa K\gg R_d$ the diffusion time reaches an asymptotic value of
\begin{equation}
t_d\to\frac{\kappa K}{c}\left[\ln\left(\frac{c}{v_{\rm sh}}\right)-1\right]\sim\frac{\kappa K}{c}\ln\left(\frac{c}{v_{\rm sh}}\right).
\end{equation}

The expression $t\sim\kappa K/c$, sometimes with the logarithmic correction mentioned, was used in previous works \citep{Ofek2010,Balberg2011,ChevalierIrwin2011} to estimate the order of magnitude of the light curve time scale. Indeed, for $K\sim 10^{18}\textrm{ g cm}^{-1}$ (see \S~\ref{sec:Progenitors}), this time scale is $\sim 100$ days, which is consistent with the time scale in the relevant observations \citep{Quimby2007,Quimby2011,Pastorello2010,Smith2010}. However, since we are interested in the regime where $R_d$ and $R_w$ are of the same order of magnitude, a more careful analysis is necessary and we must take into account the dependence of $t_d$ on $R_w$, which is evident in equation \eqref{eq:tdiff_both}.
Note that the parameters chosen in \S~\ref{sec:Progenitors} do dictate that $R_w\ll\kappa K$, which means, according to equation \eqref{eq:r_ph}, that the radius of the photosphere $R_{\rm ph}$ is roughly the wind radius $R_w$. Thus, our choice of $R_w\sim 10^{15}$ cm, motivated by the observed photosphere radius (see \S~\ref{sec:Introduction}) is self consistent.

In general, the shock velocity changes with radius, and we can roughly estimate that in the regime $M_w\sim M_{\rm ej}$
\begin{equation}\label{eq:vsh_ansatz}
v_{\rm sh}\propto\left(\frac{E}{M_{\rm ej}+M_{\rm sh}}\right)^{1/2}\approx\left(\frac{E}{M_*+4\pi KR_{\rm sh}}\right)^{1/2},
\end{equation}
with $M_{\rm sh}\propto KR_{\rm sh}$ is the accumulated mass enclosed by $R_{\rm sh}$.
When $v_{\rm sh}$ is a function of $R_{\rm sh}$, and therefore a function of $R_w$ the functional dependence of $t_d$ on $R_w$ can be more complex than presented in equation \eqref{eq:tdiff_both}. Nonetheless, we can gain significant insight by relateing $v_{\rm sh}$ to the progenitor system parameters in the limits where self-similar solutions exist. The fundamental point is that for a fixed value of $K$, the limit $R_w\gg R_d$ also corresponds to $M_w\gg M_{\rm ej}$, while in the opposite case of 
$R_w\ll R_d$ we also have a low mass wind, $M_w\ll M_{\rm ej}$. 

In a very massive wind dimensional analysis similar to the Sedov-Taylor problem \citep[see][]{Zeldovich} can be applied, leading to a shock radius and shock velocity which evolve as  
\begin{equation}\label{eq:sedov_rsh}
R_{\rm sh}\propto\left(\frac{Et^2}{K}\right)^{1/3} ;\; v_{\rm sh}=\dot{R_{\rm sh}}\propto\left(\frac{E}{KR_{\rm sh}}\right)^{1/2}.
\end{equation} 
However, since in this limit breakout occurs at $R_{\rm sh}\approx R_d$, independent of $R_w$, the shock velocity at breakout is also independent of $R_w$. Equation \eqref{eq:tdiff_both} provides a qualitative understanding of the dependence of $t_d$ on $K$ and $E$; neglecting the logarithmic correction, we have (for $R_w\gg R_d$)
\begin{equation}\label{eq:tdiff_ke}
t_d\propto K^1E^0.
\end{equation}
  
For a very low mass wind, $ v_{\rm sh}$ can be found with the alternative self-similar solution of \citet{Chevalier1982} who studied the interaction of ejecta, with the unshocked density profile  $\rho_{\rm ej}(r,t)\propto r^{-m}t^{m-3}$, and wind with density profile $\rho_w(r)\propto r^{-s}$. In our case $s=2$ and $m$ can be related \citep[see][]{Matzner1999,Rabinak2011} to the polytropic index $n$: $m\approx 10$ (for $n=3$) or $m\approx 12$ (for $n=3/2$). The shock propagates in time with radius
\begin{equation}\label{eq:rsh_chev}
R_{\rm sh}\propto E^{\frac{1}{2}(m-3)/(m-s)}K^{-1/(m-s)}t^{(m-3)/(m-s)},
\end{equation}
and with velocity given by
\begin{equation}\label{eq:vsh_chev}
v_{\rm sh}=\dot{R_{\rm sh}}\propto E^{1/2}K^{-1/(m-3)}R_{\rm sh}^{(s-3)/(m-3)}.
\end{equation}
Combining equation \eqref{eq:vsh_chev} with equation \eqref{eq:tdiff_both} results in an inverted dependence of the diffusion time on $K$ in the regime $R_w\ll R_d$: 
\begin{equation}\label{eq:tdiff_k_chev}
t_d\propto E^{-1}K^{2/(m-3)-1}\approx
\begin{cases}
E^{-1}K^{-5/7} & m\approx 10~(n=3) \\
E^{-1}K^{-7/9} & m\approx 12~(n=3/2)
\end{cases}.
\end{equation}  
In this regime $R_{\rm sh}\approx R_w$ and, using equation \eqref{eq:vsh_chev}, we have
\begin{equation}\label{eq:vsh_small}
v_{\rm sh}\propto R_w^{-1/(m-3)}\approx
\begin{cases}
R_w^{-1/7} & m\approx 10~(n=3) \\
R_w^{-1/9} & m\approx 12~(n=3/2)
\end{cases}.
\end{equation}
This is a weak dependence on $R_w$, which would result in a small deviation from the relation $t_d\propto R_w^2$ of equation \eqref{eq:tdiff_both}.
The inversion of the dependence of $t_d$ on $K$ between limits (linear in $K$ for $R_w\gg R_d$ and inverse for $R_w\ll R_d$) suggests a weak dependence on $K$ for the intermediate regime.
We note that, in theory, if the Sedov-Taylor solution were applied to the $R_w\ll R_d$ case, by combining equation \eqref{eq:sedov_rsh} with equation \eqref{eq:tdiff_both}, we would also have the $t_d\sim E^{-1}$ dependence, implying the robustness of this result.
 
We conclude that when identifying the limits of $R_w\gg R_d$ and $R_w\ll R_d$ with $M_w\gg M_{\rm ej}$ and $M_w\ll M_{\rm ej}$, respectively, then at both limits $v_{\rm sh}$ is approximately constant (a different constant for each limit), justifying equation \eqref{eq:tdiff_both}. We expect the region $M_w\sim M_{\rm ej}$ to exhibit some deviation from these time scale estimates.

\section{The Code}
\label{sec:the_code}

We have written a 1D Lagrangian computer program in order to calculate the shock propagation and light curve. In this section we describe the code in brief. The code uses the standard von Neumann and Richtmyer staggered mesh method \citep{vonNeumann,Richtmyer} to solve the nonrelativistic equations of motion. The energy equation is solved implicitly, and the radiative flux is added to the hydrodynamics assuming local thermal equilibrium (LTE) and in the diffusion approximation \citep[see][]{Zeldovich}. More specifically, the code solves the following energy equation (implicitly, solving a tridiagonal equation system):
\begin{subequations}
\begin{align}
&\frac{\partial e}{\partial t}+p\frac{\partial V}{\partial t}+\frac{1}{\rho}\nabla F=0 \\
&F=-D\nabla \left(aT^4\right) \label{eq:code_flux}
\end{align} 
\end{subequations}
where $e$ is the specific energy, $V=1/\rho$ the specific volume and $p$ is the pressure. $F$ denotes the radiative flux, with $D=\lambda c/3$ as the diffusion coefficient, and $a$ the radiation constant. The temperature $T$ in equation \eqref{eq:code_flux} is the temperature of the fluid (a result of the LTE assumption). In the case of constant opacity, the mean free path satisfies $\lambda=1/\kappa\rho$.
Gravitation can be neglected for the description of the shock propagation and breakout, since $GM_*^2/R_*\ll E$.

The equation of state (EOS) is that of a perfect gas, with radiation terms added to the pressure and energy:
\begin{align}
&p(\rho,T)=\Gamma\rho T+\frac{aT^4}{3} \\
&e(\rho,T)=\frac{\Gamma T}{\gamma-1}+\frac{aT^4}{\rho}
\end{align}
where $\Gamma=R/\mu$ is the gas constant divided by the molar mass. In the examples shown in this section we choose $\gamma=5/3$, suitable for monoatomic gas and $\mu=0.6$ which corresponds to a fully ionized mixture of hydrogen and helium with primordial ratios. In the context of shock breakout the energy and pressure of the fluid are dominated by radiation after the shock passage and during the period of adiabatic expansion which follows, so these terms have a minimal effect on the equation of state and the light curve.

The assumption of thermal equilibrium and the validity of the diffusion approximation are reasonable for nonrelativisitc shocks \citep{Katz2010}, and we rely on them here for our estimates. We do note that near the photosphere during shock breakout \citep{Sari2010,Rabinak2011} transport, rather than diffusion, is a more appropriate description of the photon propagation, which we will apply in future work.

As a code check, we calculated several test problems with our program and compared the results to known solutions. Among the test problems we considered are the self similar interaction of ejecta and wind \citep{Chevalier1982}, Elliot's extension to the Sedov-Taylor explosion which includes radiative flux \citep{Elliot1960}, and planar shock breakout \citep{Sapir2011}. We present our results for the planar shock breakout in the Appendix.

\begin{figure}[h]
\epsscale{1} \plotone{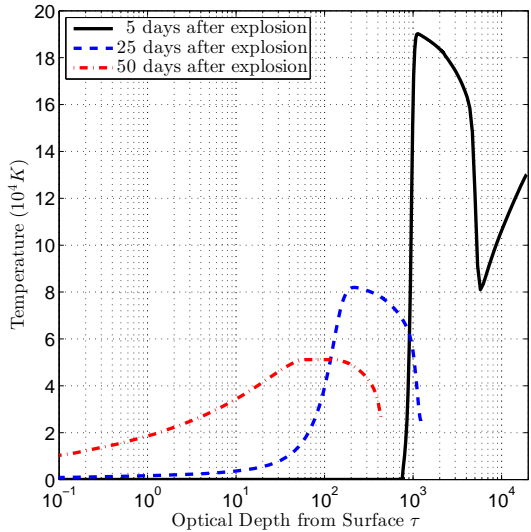}
\caption{Calculated temperature profiles at different times following explosion. The profiles were obtained with the parameters $R_w=2.5\times 10^{15}$ cm, $K=10^{18}\textrm{ g cm}^{-1}$, and $E=5\times 10^{51}$ erg. The plotted profiles are from 5 (solid black line), 25 (dashed blue line), and 50 (dot-dashed red line) days following the explosion.
\label{fig:tmp_prof}}
\end{figure}

\begin{figure}[h]
\epsscale{1} \plotone{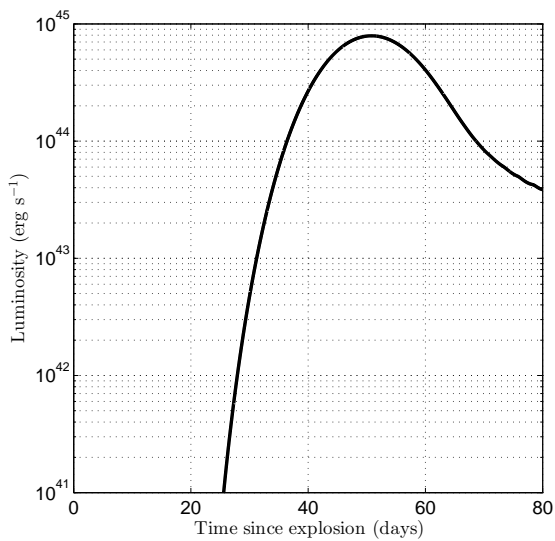}
\caption{Calculated light curve. The light curve was calculated with the same model as figure \ref{fig:tmp_prof}.
\label{fig:tmp_prof_curve}}
\end{figure}

We present an example of a calculation in figures \ref{fig:tmp_prof} and \ref{fig:tmp_prof_curve}. Figure \ref{fig:tmp_prof} shows the temperature profiles at different times. The transition from a discrete shock front to breakout at $\tau\sim c/v_{\rm sh}$, as discussed in \S~\ref{sec:QualitativePicture}, is evident in the late time (50 days after explosion) profile. In figure \ref{fig:tmp_prof_curve} the corresponding light curve is presented. It can be seen that the emergence of a light curve and its duration correspond to the shock breakout time scales. 

\subsection{Comparison with \citet{Moriya2011}}
\label{sec:moriya}

\begin{figure}[h]
\epsscale{1} \plotone{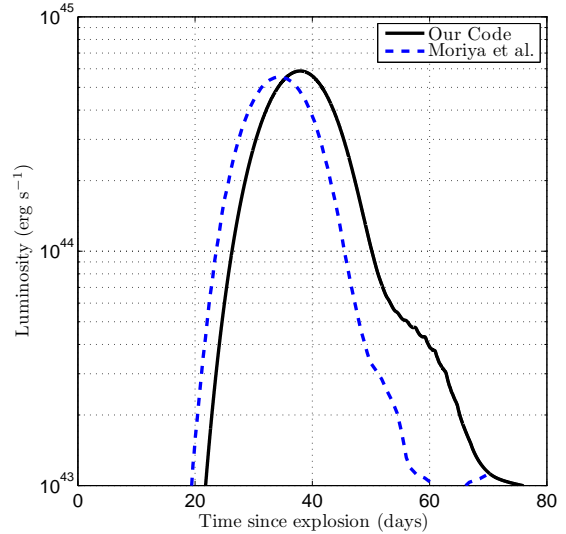}
\caption{Comparison of the light curve calculated by our code (solid black line) to the one calculated by \citet{Moriya2011} for the model s15w2r20m1e3 (dashed blue line). For the comparison we used the same progenitor profile as \citet{Moriya2011}.
\label{fig:moriya}}
\end{figure}

As a final test, we compared our numerical results with those of \citet{Moriya2011} who also studied the effect of a wind on light curves. Their work focused on less dense winds and using a more complex numerical model which we do not try to reproduce. However, we can compare our calculations to the progenitor-wind model which is the most relevant to the regime studied in the current work:  model s15w2r20m1e3, marked as $10^{-1}M_\sun \textrm{yr}^{-1}$ in figures 3,4 of \citet{Moriya2011}, which has a wind mass of $6.5M_\sun$. In figure \ref{fig:moriya} we show a comparison between the light curve calculated with our code and their results. Our light curve was calculated using the progenitor profile given in figure 3 of \citet{Moriya2011}, which is slightly different from our standard $M_*=15M_\sun$ progenitor (see \S~\ref{sec:Progenitors}). The resulting relevant light curve parameters (peak luminosity, time duration, and total radiated energy) are the same up to few percents. 

\section{Numerical Parameter Survey}
\label{sec:numerical_results}

In this section we study the light curves calculated with the code and conduct a parameter survey in order to map the dependence on the important parameters. We also verify that the analytical limits presented in \S~\ref{sec:QualitativePicture} are recovered, and use them to gain insight to the results.

\subsection{Comparison with analytical limits}
\label{sec:analytic}

Following a convergence test, we model the compact star part of the progenitor system with $n_*=50$ cells, geometrically decreasing in size $\Delta r$ toward the outer edge with a quotient $q_*=0.98$. The wind is divided into $n_w=250$ cells, geometrically increasing in size (with a constant quotient $q_w>1$) toward the outer wind edge. The innermost wind cell is of the same size as the outermost star cell. The size of the innermost wind cell, together with $n_w$, determine $q_w$. In these calculations we use the smoothing parameters $\alpha=1.25$ and $k=15$ (see \S~\ref{sec:Progenitors}). Generally, we set $R_*=10^{13}\;$cm, except for very low values of $R_w$, for which we had to reduce the star radius, $R_*$, in order to remain in the regime $R_*\ll R_w$. Additionally, we had to change the resolution for convergence and to disable the wind edge smoothing (see \S~\ref{sec:Progenitors}). These changes have an effect only at very low $R_w$. 

\begin{figure}[h]
\epsscale{1} \plotone{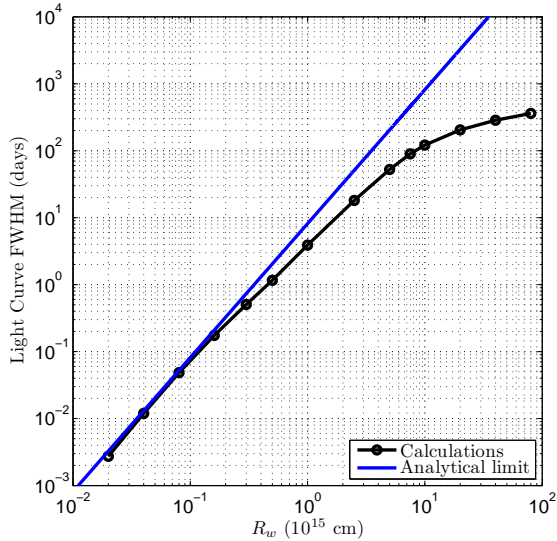}
\caption{Full Width at Half Maximum (FWHM) of the calculated light curve as a function of the wind outer radius $R_w$. Results are plotted for nominal values of $K=10^{18}\textrm{ g cm}^{-1}$ and $E=5\times 10^{51}\textrm{ erg}$ (black line, marked with circles). Each marker represents a single hydrodynamic diffusion calculation. The analytical limit (solid blue line), following equation \eqref{eq:tdiff_both}, is of the form $t_d=R_w^2\left(\kappa K/c\right)^{-1}v_{\rm sh}^{-2}$, with fitted parameter $v_{\rm sh}=3.5\times 10^8\textrm{ cm s}^{-1}$
\label{fig:range_small}}
\end{figure}

\begin{figure}[h]
\epsscale{1} \plotone{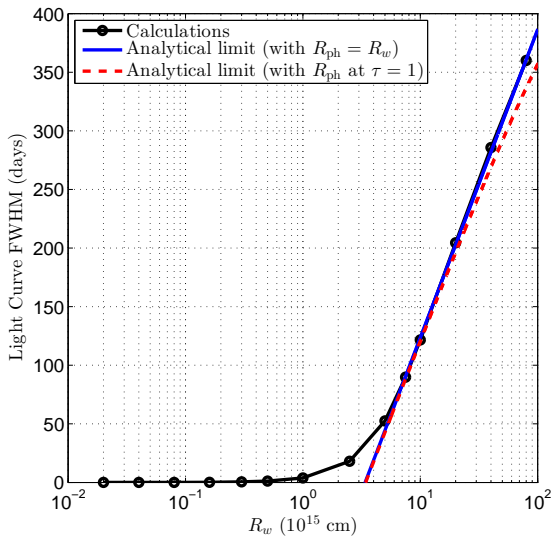}
\caption{Full Width at Half Maximum (FWHM) of the calculated light curve as a function of the wind outer radius $R_w$. Results are plotted for nominal values of $K=10^{18}\textrm{ g cm}^{-1}$ and $E=5\times 10^{51}\textrm{ erg}$ (black line, marked with circles). Each marker represents a single hydrodynamic diffusion calculation. The calculations are the same as in figure \ref{fig:range_small}, but plotted in a different scale. The analytical limit (solid blue line), following equation \eqref{eq:td_big_approx} for the approximation $R_{\rm ph}=R_w$ (which is used by the code; see text), is of the form $t_d=\eta(\kappa K/c)[\ln(R_w/R_d)-1]$, with fitted parameters $\eta=0.9$ and $R_d=1.2\times 10^{15}$ cm (or equivalently $v_{\rm sh}=1.1\times 10^8 \textrm{ cm s}^{-1}$). A second, more adequate, analytical limit (dashed red line), with the same $v_{\rm sh}$ and multiplication factor $\eta$, is in the form of equation \eqref{eq:tdiff_both}. This limit takes into account the photosphere at $\tau=1$. 
\label{fig:range_big}}
\end{figure}

We relate the diffusion time scale $t_d$ with the full width at half maximum (FWHM) of the light curve. This choice is independent of the low luminosity "tail" at late times, which arises in part from the continued interaction of the shock wave and the wind \citep{ChevalierIrwin2011}, and is sensitive to the shape of the wind cutoff profile. The results, for $E=5\times 10^{51}$ erg, $K=10^{18}\textrm{ g cm}^{-1}$, and a wide range of $R_w$ are shown in figures \ref{fig:range_small} and \ref{fig:range_big}.

Both figures demonstrate that the calculated light curves do indeed reproduce the analytic timescales at the appropriate limits. The fit is very good at low values of $R_w$, while at high values there is some deviation from equation \eqref{eq:tdiff_both}. The reason is that the simplified treatment of diffusion in the code extends all the way to $R_w$, whereas our analytic estimates were based on emission from a $\tau=1$ (or $\tau=2/3$) surface. A more exact fit of the numerical results is therefore found with a revised analytical estimate of the diffusion time using equation \eqref{eq:tdiff_integrated}, and substituting $R_{\rm ph}=R_w$:
\begin{equation}\label{eq:td_big_approx}
t_d\approx\frac{\kappa K}{c}\left[\ln\frac{R_w}{R_d}-1\right]=\frac{\kappa K}{c}\left[\ln\left(\frac{c}{v_{\rm sh}}\frac{R_w}{\kappa K}\right)-1\right].
\end{equation} 
Unlike equation \eqref{eq:tdiff_both}, equation \eqref{eq:td_big_approx} does not reach an asymptotic value at $R_w\gg\kappa K$. Correspondingly, our numerical calculations which do not take into account the photosphere at $\tau\sim 1$ are inexact in this sense. However, since we deal with the regime $R_w\ll\kappa K$, the difference between equations \eqref{eq:tdiff_both} and  \eqref{eq:td_big_approx} is small (see figure \ref{fig:range_big}, which even approaches $R_w\sim\kappa K$), so the numerical treatment of the photosphere is sufficient for our purposes in the current work.  

\subsection{Progenitor system - light curve relation}
\label{sec:relation}

\begin{figure}[h]
\epsscale{1} \plotone{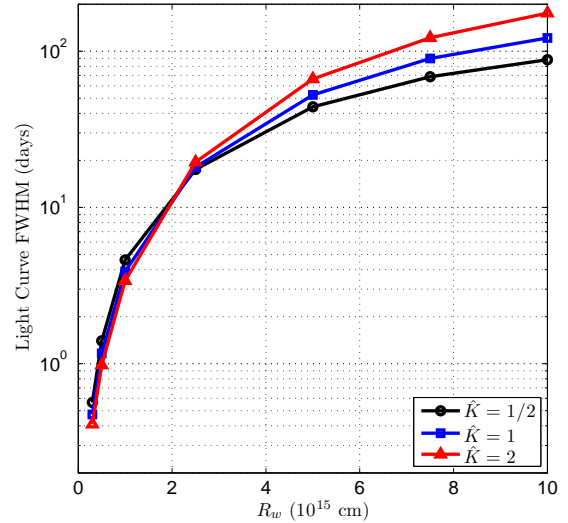}
\caption{Full Width at Half Maximum (FWHM) of the calculated light curve as a function of the wind outer radius $R_w$. Results are plotted for nominal values  $\hat{K}\equiv K/\left(10^{18}\textrm{ g cm}^{-1}\right)=\hat{E}\equiv E/\left(5\times 10^{51}\textrm{ erg}\right)=1$ (blue line, marked with squares), for $\hat{K}=1/2$ (black line, marked with circles), and for $\hat{K}=2$ (red line, marked with triangles). Each marker represents a single hydrodynamic diffusion calculation. 
\label{fig:time_k}}
\end{figure}

\begin{figure}[h]
\epsscale{1} \plotone{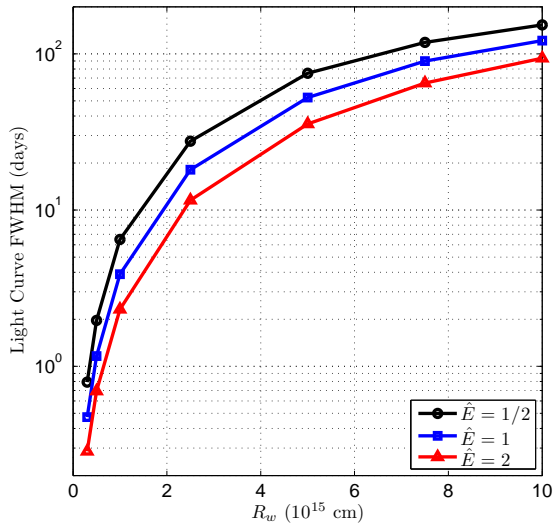}
\caption{Full Width at Half Maximum (FWHM) of the calculated light curve as a function of the wind outer radius $R_w$. Results are plotted for nominal values  $\hat{K}\equiv K/\left(10^{18}\textrm{ g cm}^{-1}\right)=\hat{E}\equiv E/\left(5\times 10^{51}\textrm{ erg}\right)=1$ (blue line, marked with squares), for $\hat{E}=1/2$ (black line, marked with circles), and for $\hat{E}=2$ (red line, marked with triangles). Each marker represents a single hydrodynamic diffusion calculation. 
\label{fig:time_e}}
\end{figure}

As is obvious from the qualitative discussion in \S~\ref{sec:QualitativePicture} and from figures \ref{fig:range_small} and \ref{fig:range_big}, the range of $R_w$ of interest in this work ($R_w\sim 10^{15}$ cm) cannot fit well to any of the analytical limits. Henceforth we conduct a numerical parameter survey in this range. We begin by examining the relation between the diffusion time, $t_d$, and the parameters of the progenitor system, namely $E$,$K$ and $R_w$, whereas for simplicity we fix 
$M_*=15M_\sun$. We study the dependence of $t_d$ on $R_w$ for a nominal calculation with $E=E_0\equiv 5\times 10^{51}$ erg and $K=K_0\equiv 10^{18}\textrm{ g cm}^{-1}$. The dependence on $E$ and $K$ is studied by repeating the calculations with other values of $\hat E\equiv E/E_0$ and $\hat K\equiv K/K_0$. The results are shown in figures \ref{fig:time_k} and \ref{fig:time_e}.

The increase of $t_d$ with increasing $R_w$ is understood qualitatively by equation \eqref{eq:tdiff_both}. Quantitatively, we see that $t_d$ is strongly dependent on $R_w$ throughout the relevant range, so the approximation $t_d\approx\kappa K/c$ \citep{Ofek2010,Balberg2011,ChevalierIrwin2011} can serve only as an order of magnitude estimate. This behavior is also evident in figure 9 of \citet{Moriya2011}. As can be seen in Figure \ref{fig:time_k}, the dependence of the diffusion time on $K$ is in good agreement with the analysis of \S~\ref{subsec:Ltdiff}. At large $R_w$, the diffusion time increases roughly linearly with $K$, but for lower $R_w$ the dependence becomes weaker and is finally inverted, as expected in low mass wind (see equations \eqref{eq:tdiff_ke}, \eqref{eq:tdiff_k_chev}), yielding smaller diffusion times for larger values of $K$. 
We also recover the $t_d\sim E^{-1}$ relation expected at low $R_w$ (again, see equation \eqref{eq:tdiff_k_chev}). The dependence becomes weaker at larger $R_w$, as might be expected from equation \eqref{eq:tdiff_ke}. 

\begin{figure}[h]
\epsscale{1} \plotone{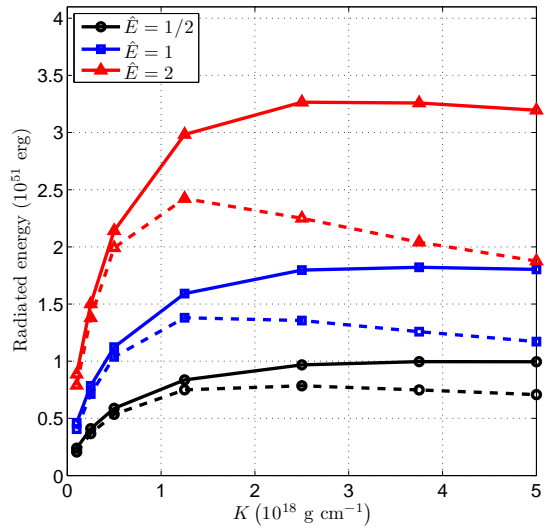}
\caption{Radiated energy as a function of the wind density coefficient $K$, for winds with outer radius $R_w=2.5\times 10^{15}$ cm. Results are plotted for a nominal value $\hat{E}\equiv E/\left(5\times 10^{51}\textrm{ erg}\right)=1$ (blue lines, marked with squares), for $\hat{E}=1/2$ (black lines, marked with circles), and for $\hat{E}=2$ (red lines, marked with triangles). Each marker represents a single hydrodynamic diffusion calculation. For each value of $\hat{E}$ the total radiated energy is plotted as a solid line and a Gaussian estimate for the radiated energy, based on the peak luminosity and FWHM of the light curve (see text), is plotted as a dashed line.  
\label{fig:energy_lum}}
\end{figure}

The other feature in the light curve which we relate to the parameters of the progenitor system is the total radiated energy. Figure \ref{fig:energy_lum} shows the dependence of the radiated energy on $K$ for different values of $\hat{E}$ and for a fixed $R_w=2.5\times 10^{15}$ cm. We calculate the radiated energy by integrating the light curve until the luminosity drops to 0.1\% of the maximum luminosity.  The behavior exhibited in figure \ref{fig:energy_lum} is qualitatively understood by the plastic collision relation (equation \eqref{eq:e_rad}), thereby saturating when $M_w\gtrsim M_*$. Using \eqref{eq:wind_mass1}, $M_w\approx M_*$ for $\hat{K}=1$. 

We digress and discuss the shape of the light curve because the total radiated energy is not always reliably observed. When observations are limited to the vicinity of peak magnitude and do not track low luminosities, the shape of the light curve must be modeled theoretically to asses its total energy. One such model uses the observed peak luminosity $L_{\textrm{max}}$ and the observed FWHM of the light curve, and, assuming a Gaussian light curve \citep{Arnett1996}, $E_{\textrm{rad}}\approx \left(0.5\sqrt{\pi/\ln{2}}\right) L_{\textrm{max}}\times\textrm{FWHM}$. We plot this estimate of the radiated energy (using the calculated peak luminosity and the FWHM of the calculated light curve) as dashed lines in figure \ref{fig:energy_lum}. It is obvious from figure \ref{fig:energy_lum} that as the wind becomes more dense, this simple estimate deviates more from the total radiated energy (compare the solid and dashed lines in figure \ref{fig:energy_lum}). The reason is that the late time "tail" strongly deviates from a Gaussian form, and includes much more energy than predicted by a Gaussian approximation. This tail is partially powered by continued interaction of the shock wave and wind \citep{ChevalierIrwin2011} and thus becomes more prominent and contains more energy. We can conclude that total energy estimates based on observations should be done carefully, and that a Gaussian fit must be treated as a lower limit if only the vicinity of the peak region of the light curve is observed. 

\subsection{Time scale constraints}
\label{sec:time_scale}

\begin{figure}[h]
\epsscale{1} \plotone{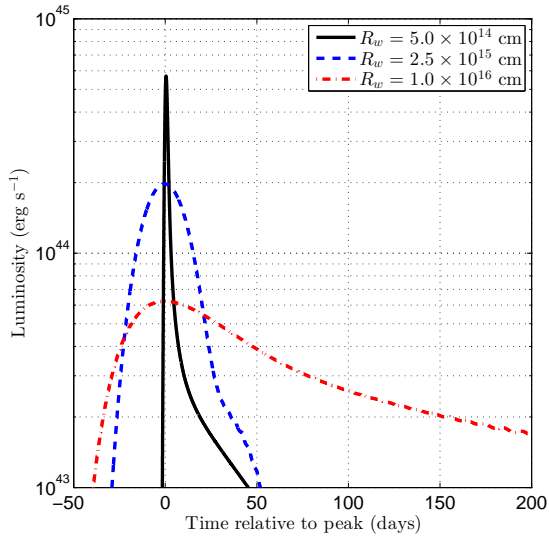}
\caption{Calculated light curves for different wind radii. The plotted light curves are obtained from models with the same released energy $E=2\times 10^{51}$ erg and the same wind mass (obtained by keeping $KR_w=2.5\times 10^{33}$ g constant). The plotted light curves have wind radii of $R_w=5.0\times 10^{14}$ cm (solid black line), $R_w=2.5\times 10^{15}$ cm (dashed blue line), and $R_w=1.0\times 10^{16}$ cm (dot-dashed red line), corresponding to density coefficients values of $K=5\times 10^{18}\textrm{ g cm}^{-1}$, $K=1.0\times 10^{18}\textrm{ g cm}^{-1}$, and $K=2.5\times 10^{17}\textrm{ g cm}^{-1}$ respectively.
\label{fig:2010timescale}}
\end{figure}

In this work we focus on winds with radius $R_w\sim 10^{15}$ cm. The justification we gave for this choice is the observed photosphere radius $R_{\rm ph}\sim 10^{15}$ cm and the relation $R_w\sim R_{\rm ph}$ which is valid for $M_w\sim M_{\rm ej}$ (see \S~\ref{sec:QualitativePicture}). One of the conclusions of the numerical results is that the observed light curve time scale gives another constrain on $R_w$. We demonstrate this point in Figure \ref{fig:2010timescale}, where we show the light curves  calculated for three combinations of $K$ and $R_w$, all of which satisfy $K R_w=2.5\times 10^{33}$ g, so that the total wind mass is kept constant. The intermediate model with $R_w=2.5\times 10^{15}$ cm is used below to fit SN 2010gx (see \S~\ref{sec:observations}), while the other two are more compact and extended winds (note that for the extended wind, the approximation $R_w\ll\kappa K$ which our code assumes is marginal). By keeping the total wind mass constant the efficiency of converting a given explosion energy to radiated energy is fixed (for a given progenitor star mass) and so the light curve time scale becomes a direct indicator of the wind radius. It is obvious that for time scales of $\sim 50$ days, as is the case of SN 2010gx, the models with the more compact and more extended winds are ruled out.   

\section{Observations}
\label{sec:observations}

In this section we relate our model to SLSNe observations. Specifically, we focus on three events: SN 2010gx, SN 2006gy and SN 2005ap. Our goal is to ascertain that a steady wind model can provide a viable explanation for the observed SLSNe light curves, and to constrain the likely parameters of the progenitor system. In general, previous works which considered a star-wind system \citep{SmithMcCray2007,Smith2010,ChevalierIrwin2011} provided only order of magnitude correlation between the steady wind model and the observations. Very recently, \citet{Moriya2012new} have presented a specific numerical model for SN 2006gy where the ejecta interacts with a a distant shell (rather than a steady wind), and we comment on the similarities and differences regarding this specific object below. 

\subsection{SN 2010gx}
\label{sec:sn2010gx}

\begin{figure}[h]
\epsscale{1} \plotone{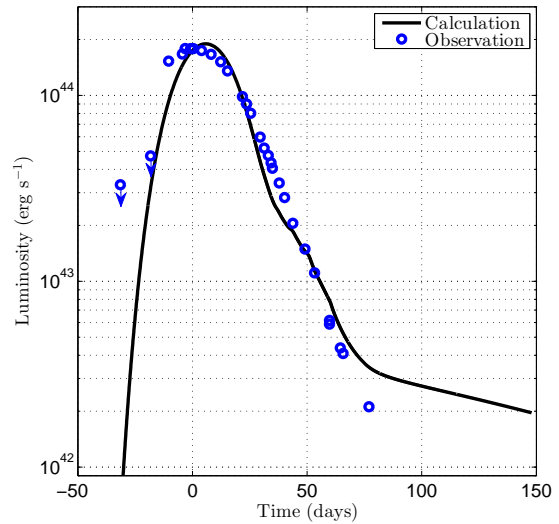}
\caption{Calculated and observed light curves of SN 2010gx. The calculated light curve (solid black line) was calculated with the parameters $R_w=2.5\times 10^{15}$ cm, $K=10^{18}\textrm{ g cm}^{-1}$, and $E=2\times 10^{51}$ erg. The observed light curve (blue circles) is taken from \citet{Pastorello2010}, with zero bolometric correction.   
\label{fig:sn2010gx}}
\end{figure}

We take the data for SN 2010gx from \citet{Pastorello2010}. The measured luminosity and temperature imply a blackbody radius $\sim 3\times 10^{15}$ cm at peak luminosity. Since the photosphere lies close to the wind edge (see \S~\ref{sec:QualitativePicture}), we choose a model with $R_w=2.5\times 10^{15}$ cm (the photosphere expands following the passage of the shock). We find that the light curve is recovered well when setting the other parameters to be $K=10^{18}\textrm{ g cm}^{-1}$ and $E=2\times 10^{51}$ erg. The quality of the fit is shown in figure \ref{fig:sn2010gx}, which compares the calculated and observed light curves. It is noteworthy that we do include a correction for the difference in arrival times of photons originating from different positions on the photosphere \citep{Katz2012}. This effect on the time scales is $\sim R_{\rm ph}/c$, which is about a day, and therefore negligible in the cases considered here. 

\begin{figure}[h]
\epsscale{1} \plotone{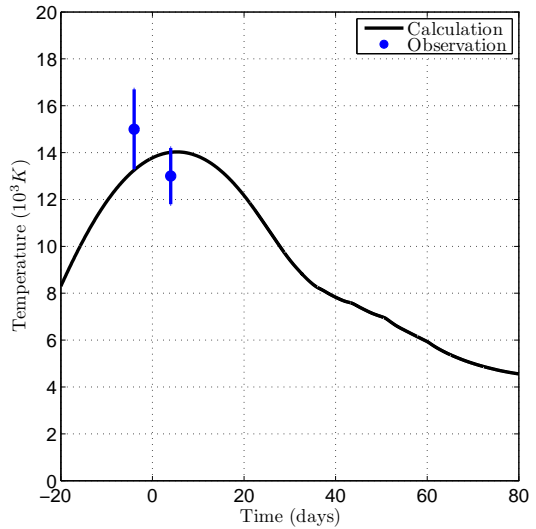}
\caption{Calculated and observed blackbody temperature of SN 2010gx. The calculated temperature (solid black line) was calculated with the same model as in figure \ref{fig:sn2010gx}. The observed temperature (blue circles with error-bars) is taken from \citet{Pastorello2010}. 
\label{fig:sn2010gx_tmp}}
\end{figure}

The corresponding blackbody temperature for the calculated light curve is compared with temperatures inferred from observations in figure \ref{fig:sn2010gx_tmp}. We emphasize that the calculated blackbody temperatures are recovered using the luminosity, $L(t)$, and identifying the photosphere, $R_{\rm ph}(t)$, with the position where the optical depth is $\tau=2/3$, 
\begin{equation}
\label{eq:black_body}
L=4\pi R_{\rm ph}^2\sigma T^4.
\end{equation}
Clearly this determination of the blackbody temperature is a crude one, and, in fact, the blackbody assumption must generally be considered only as an approximation \citep{Sari2010,Rabinak2011,ChevalierIrwin2011}. Therefore, we view the fit between calculated and observed temperatures as indicative that our model is compatible with observations.. 

Considering the simplifications of our approach, we do not attempt to find a "best fit" model for the progenitor system, but rather demonstrate that a plausible range exists. Moreover, we emphasize that the wind parameters we choose to fit are not unique. If we fix $R_w=2.5\times 10^{15}$ cm, as implied by the temperature measurement, we have a considerable degree of freedom to choose $K$. The reason is the weak dependence of the diffusion time on $K$ for this $R_w$, as seen in figure \ref{fig:time_k}. The total radiated energy is also not strongly effected by an increase of $K$, as seen in figure \ref{fig:energy_lum}. This is because $M_w\approx 4\pi KR_w$ is close to the saturation region of equation \eqref{eq:e_rad} and figure \ref{fig:energy_lum}. More specifically, a different model with  $K=2\times 10^{18}\textrm{ g cm}^{-1}$ (double the chosen value), results in a light curve with the same peak luminosity as the model in figure \ref{fig:sn2010gx} and FWHM wider by only 15\%. In addition, we tested a model with $\kappa=0.2\textrm{ cm}^2\textrm{g}^{-1}$, which is appropriate for a hydrogen poor CSM, as indicated by SN 2010gx observations \citep{Pastorello2010}. This change has a small (less than 10\%) effect on the light curve and a negligible effect on the temperature, due to the change from inverse to linear dependence on $\kappa$ in equation \eqref{eq:tdiff_both} (which results in a weak dependence at $R_w\sim R_d$).

\subsection{SN 2006gy}
\label{sec:sn2006gy}

\begin{figure}[h]
\epsscale{1} \plotone{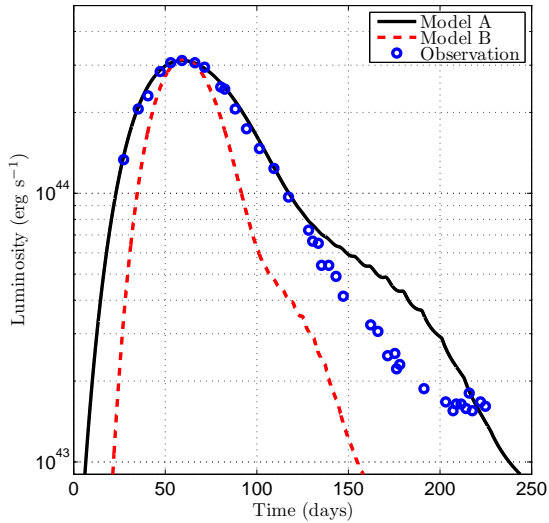}
\caption{Calculated and observed light curves of SN 2006gy. Model A (solid black line) was calculated with the parameters $R_w=6.5\times 10^{15}$ cm, $K=0.9\times 10^{18}\textrm{ g cm}^{-1}$, and $E=5\times 10^{51}$ erg. Model B (dashed red line) was calculated with the parameters $R_w=4\times 10^{15}$ cm, $K=10^{18}\textrm{ g cm}^{-1}$, and $E=3.7\times 10^{51}$ erg. The observed light curve (blue circles) is taken from \citet{Smith2010}, with bolometric correction.     
\label{fig:sn2006gy}}
\end{figure}

We take the data for SN 2006gy from \citet{Smith2010}. The measured luminosity and temperature imply a blackbody radius $\sim 4.5\times 10^{15}$ cm at peak luminosity. If we do not consider the temperature constraint, a model which fits the light curve can be found. The model with $R_w=6.5\times 10^{15}$ cm, $K=0.9\times 10^{18}\textrm{ g cm}^{-1}$, and $E=5\times 10^{51}$ erg, which reasonably fits the light curve except for the tail, is plotted in figure \ref{fig:sn2006gy} (solid black line). We shall distinguish this model as model A.  The wind mass in this model is $M_w\approx 35M_\sun$, which is similar to the estimates of previous works \citep{Woosley2007,Smith2010,ChevalierIrwin2011}. 

\begin{figure}[h]
\epsscale{1} \plotone{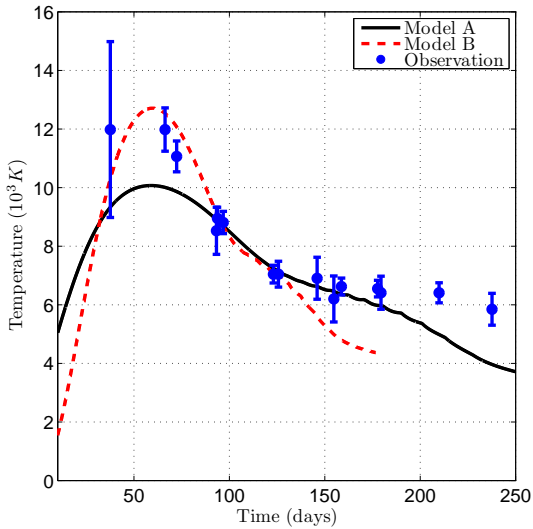}
\caption{Calculated and observed blackbody temperature of SN 2006gy. The models (A - solid black line, B - dashed red line) were calculated with the same parameters as in figure \ref{fig:sn2006gy}. The observed temperature (blue circles with error-bars) is taken from \citet{Smith2010}
\label{fig:sn2006gy_tmp}}
\end{figure}

Model A may fit the light curve, but since $R_w$ in this model is larger than the implied blackbody radius, it deviates from the temperature measurements, as can bee seen in figure \ref{fig:sn2006gy_tmp}. We find that a model in a different range of the progenitor parameters can be chosen to reproduce the observed temperatures (assuming a blackbody emission) and the peak luminosity, but at the expense of generating a light curve which is too narrow. One such model, distinguished as model B, is also plotted in figures \ref{fig:sn2006gy} and \ref{fig:sn2006gy_tmp}. In this model we set $R_w=4\times 10^{15}$ cm, $K=10^{18}\textrm{ g cm}^{-1}$, and $E=3.7\times 10^{51}$ erg. 

The light curve is narrow due to the strong dependence of the diffusion time on $R_w$, which allows little freedom (see \S~\ref{sec:time_scale}). As a result, we cannot find a single model which fits well both the light curve and the measured temperature. This is easily understood by considering the analysis in \S~\ref{sec:relation}. If we use the wind radius fixed by the observed peak luminosity and temperature, then for a given progenitor star only two free parameters remain: $K$ and $E$. Model B (which assumes the observationally inferred $R_w$) is adjusted to reproduce the observed peak luminosity, but it results in a narrow light curve, yielding a total radiated energy which is too low. The total radiated energy cannot be increased by increasing $K$, because at $K=10^{18}\textrm{ g cm}^{-1}$ the efficiency of converting the explosion energy is already close to the asymptotic value (see figure \ref{fig:energy_lum}; we note that in this figure $R_w=2.5\times 10^{15}$ cm; for  $R_w=4\times 10^{15}$ cm, asymptotic efficiency is reached for even smaller values of $K$, due to the larger wind mass). Stipulating a larger explosion energy $E$ can account for the total radiated energy, of course, but it is not a solution, since it leads to even narrower light curves, due to larger expansion velocities (see figure \ref{fig:time_e}). In theory, the desired effect can be obtained by increasing both the total energy and $K$ (the latter compensating for the narrowing of the light curve, since it generates a more massive wind). However, the dependence of the diffusion time on $K$ is weak for $R_w$ in the range of interest (see figure \ref{fig:time_k}), and we determine that in order to reproduce the observations, an unrealistically heavy wind mass (hundreds of solar masses) is required.

We note that the fit with observations cannot be improved by stipulating a larger mass for the progenitor star. While an increasing $M_*$ leads to a decrease in $v_{\rm sh}$ and thus to wider light curves, it works to reduce the efficiency of converting the explosion energy to radiation.  We found that even tripling $M_*$ (together with $M_w$, to keep the efficiency), which again results in unrealistically large masses, does not widen the light curve enough. 

In essence, it appears difficult to reconcile both the observed light curve and temperature of SN 2006gy with a steady wind model, since the light curve time duration implies a wind radius $R_w$ that is different from the one inferred by the photosphere radius. The reason may be the blackbody interpretation \citep[see][]{Sari2010,Rabinak2011,ChevalierIrwin2011} which effects the calculated temperature, an inaccurate assumption of full ionization (which effects the opacity, and therefore the photosphere radius), or an indication that a different model, perhaps with a different CSM profile \citep[see][for example]{Moriya2012}, is required. 

Other, similar, models for SN 2006gy have been suggested in previous works. \citet{ChevalierIrwin2011} suggested a steady wind model with estimated parameters $R_w\sim 10^{16}$ cm, $E\sim 3\times 10^{51}$ erg and $K\sim 0.5\times 10^{18}\textrm{ g cm}^{-1}$, resulting in $M_w\sim 30 M_\sun$. These parameters are similar to the ones we adjusted. Note that the large $R_w$ in this model results in a low blackbody temperature, as in our model A. 
\citet{Woosley2007} considered a pulsational pair instability scenario, where a shell of $\sim 30M_\sun$ was ejected to a radius of $\sim 10^{16}$ cm prior to the explosion. In this model, the photosphere radius remains relatively low (this is due to the shell density profile and the large radii, which allow for a lower optical depth for the same CSM mass) but apparently not low enough to fit to the temperature measurement. \citet{Chatzopoulos2012} try to explain SN 2006gy with a semi-analytical model based on the self-similar solutions of \citet{Chevalier1982}. However, their model has a low wind mass ($5M_\sun$ compared to $40M_\sun$ ejecta mass), and a small CSM radius ($2.5\times 10^{15}$ cm, albeit with a different power law), which does not allow to recover the  typical time scale in SN 2006gy. This discrepancy is noted by \citet{Moriya2012new}, who calculated the \citet{Chatzopoulos2012} model numerically. \citet{Moriya2012new} suggest their own models for SN 2006gy, which include interaction with CSM shells which are not due to steady winds, but rather a finite shell situated at a standoff distance form the star (see their models D2 and F1). The CSM in these models extends to larger radii compared to our models ($1\times 10^{16}-2\times 10^{16}$ cm), and contain less mass ($15M_\sun-18M\sun$), which is compensated by a greater ejecta energy ($10\times 10^{51}$erg). A more detailed comparison with these different models is beyond the scope of the current work, but we emphasize that our model A fits the light curve with at least the same quality as the other numerical models. One of the important aspects of \citet{Moriya2012new} is the comparison of effective and color temperatures (their figure 8). The significant difference between the two temperatures may be the reason why a simple blackbody approximation cannot account for the observed temperature.

\subsection{SN 2005ap}
\label{sec:sn2005ap}

\begin{figure}[h]
\epsscale{1} \plotone{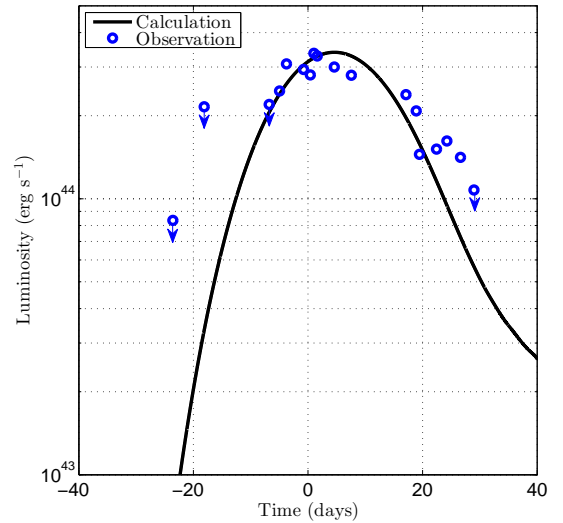}
\caption{Calculated and observed light curves of SN 2005ap. The calculated light curve (solid black line) was calculated with the parameters $R_w=2.5\times 10^{15}$ cm, $K=1.5\times 10^{18}\textrm{ g cm}^{-1}$, and $E=2.8\times 10^{51}$ erg. The observed light curve (blue circles) is taken from \citet{Pastorello2010}, with zero bolometric correction.   
\label{fig:sn2005ap}}
\end{figure}

\begin{figure}[h]
\epsscale{1} \plotone{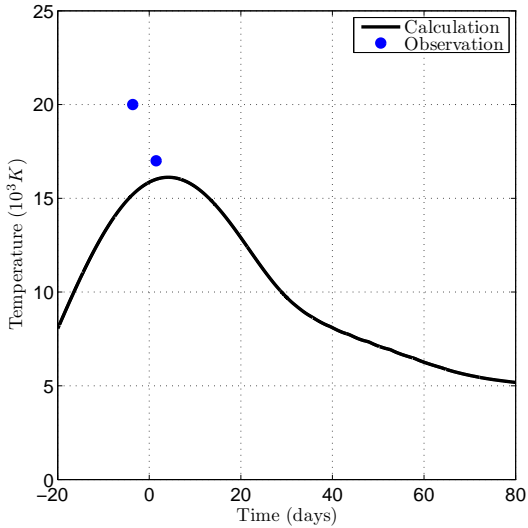}
\caption{Calculated and observed blackbody temperature of SN 2005ap. The calculated temperature (solid black line) was calculated with the same model as in figure \ref{fig:sn2005ap}. The observed temperature (blue circles) is taken from \citet{Quimby2007}. No estimate is given in \citet{Quimby2007} for the uncertainty in the two temperatures in the graph (only a general estimate of 16,000K-20,000K).
\label{fig:sn2005ap_tmp}}
\end{figure}

The light curve data for SN 2005ap is taken from \citet{Pastorello2010} and the temperature is taken from \citet{Quimby2007}. The measured luminosity and temperature imply a blackbody radius $\sim 2.5\times 10^{15}$ cm at peak luminosity. We choose a model with $R_w=2.5\times 10^{15}$ cm, $K=1.5\times 10^{18}\textrm{ g cm}^{-1}$, and $E=2.8\times 10^{51}$ erg. The calculated and observed light curves are plotted in figure \ref{fig:sn2005ap}, and  the (calculated and observed) temperatures are plotted in figure \ref{fig:sn2005ap_tmp}. The fit of the model to the observations is marginal, and suffers from problems similar to the fit to SN 2006gy, but more mildly. The model is a bit too narrow and the calculated temperature is a bit too low. This implies some discrepancy between the wind radius imposed by the diffusion time scale and the blackbody radius imposed by the temperature and maximum luminosity. We note that that as in \S~\ref{sec:sn2010gx}, the results are not sensitive to the value of $\kappa$, which is set to be smaller due to lack of hydrogen \citep[as indicated by][]{Quimby2011}. The discrepancy is again difficult to resolve by a different choice of parameters in the context of our model, but is more likely to be solved if we allow for corrections to the blackbody assumption and interpretation of observations.
As in the case of SN 2006gy, if we do not constrain $R_w$ by the implied blackbody radius (because of the non equilibrium conditions for example), it is easy to fit the light curve better by increasing $R_w$. 

\section{Conclusions and Discussion}
\label{sec:conclusions}

In this work we considered a scenario for superluminous supernovae (SLSNe) based on shock breakout from the interaction of an ejecta from an exploding star and a massive envelope of circumstellar material, presumably emitted from the star prior to the explosion. We focused on a steady wind model for this CSM, which dictates a $\rho\sim r^{-2}$ density profile, and on massive winds, which have a total mass $M_w$ comparable with the ejected mass, $M_{\rm ej}$. The latter assumption allows for shock breakout in which the entire ejecta participates, leading to an efficient conversion of the explosion energy to bolometric luminosity. This efficiency is required to explain the total energy observed in some SLSNe without resorting to high energy explosion mechanisms.

Our approach combines analytical limits and numerical hydro-diffusion calculations of the bolometric light curve. Thus we improve upon previous works which considered a massive wind in the context of SLSNe with order of magnitude estimates \citep{SmithMcCray2007,Smith2010,ChevalierIrwin2011}. We note that the numerical calculations are absolutely necessary for producing reliable results in a $M_w\approx M_{\rm ej}$ scenario, since the analytical limits cannot be applied. On the other hand, our numerical model includes only the fundamental physics of the problem, and so the principle trends are easily understood. Our calculations are in agreement with more complex numerical models \citep[see, e.g.,][]{Moriya2011}. 

Specifically, we related the main features of the observed light curve to the parameters of the progenitor system - the star and the wind. As expected, we explicitly find that a large wind mass does allow to efficiently convert the energy of the explosion into a bolometric light curve, and that this efficiency naturally saturates when $M_w > M_{ej}$. However, we demonstrate that in this scenario the width of the light curve is strongly dependent on the cutoff distance of the wind, $R_w$. Quantitatively, we show that in order to recover a time scale of tens of days seen in the most energetic SLSNe, the wind outer radius must be of order a few $10^{15}$ cm. This result is consistent with the scenario of a steady mass loss of a few $10^{-2}M_\sun\rm{ yr}^{-1}$ during the last 1000 years of the star's evolution \citep{QS2012}.

We applied our model to three SLSNe, namely SN 2010gx, SN 2006gy and SN 2005ap, which exhibited luminosities as high as several $10^{44}$ ergs with time scales of tens of days. We find that the light curves of these objects can be understood in terms of a supernova in a heavy wind scenario, where the progenitor star has a mass of $M_*\approx 15M_\sun$ and the wind mass is in the range $15M_\sun-35M_\sun$. The efficient conversion of the ejecta kinetic energy to radiation in this scenario enables us to reproduce the light curves with energies in the standard range of $2\times 10^{51}\textrm{ erg}-5\times 10^{51}$ erg, and, as mentioned above, with winds extending to radii of $2.5\times 10^{15}\textrm{ cm}-6.5\times 10^{15}$ cm. Assuming blackbody emission, our model reproduces the observations in SN 2010gx, but underestimates the temperatures measured in SN 2006gy, and to some extent in SN 2005ap as well. A similar trend was found in the other works mentioned above as well, and may be the result of a frequency dependent (rather than constant) opacity, differences between the effective and color temperatures, or both. 

We note that a separate subclass of SLSNe, where the peak luminosity  is high but the time scales are shorter, such as SN 2008es \citep{Gezari2009,Miller2009} may also be explainable with a steady wind model. In this case the wind mass must be significantly lower than the ejecta mass, so the initial wind optical depth may be small. We plan to investigate such systems numerically in future work.
  
We do not explicitly compare our results with an alternative, but similar, scenario, where the ejecta interacts with a single shell, situated at some distance from the star, presumably a result of a short period of enhanced mass loss. Clearly, a massive shell can be equally efficient in converting the explosion energy to radiation \citep{Moriya2012new}. Differentiating between both scenarios must involve a wider parameter survey (especially since in the shell model its position and density structure can be assumed independently), and probably some additional physics in the numerical model as well.

Another issue which requires further work is a non-thermal component from shock breakout through the thick wind. Recently, several authors \citep{Katz2011,ChevalierIrwin2012,Sari2012} pointed out that during breakout the shock is likely to become collisionless, hence creating a higher energy, non-thermal component in the spectrum. These works vary considerably concerning their estimates regarding the fraction of the total non-thermal energy eventually emitted from breakout, but obviously, if this fraction is sizable, there will be some impact on the bolometric light curve. 

\acknowledgements 

We are grateful to Avishay Gal-Yam, Eran Ofek, Itay Rabinak, Nir Sapir and Eli Waxman for stimulating discussions and advice regarding this work. We also wish to thank Eli Livne for helpful suggestions regarding the hydrodynamic code. 

\appendix
\section{Comparison to planar shock breakout solution}

\citet{Sakurai1960} investigated the problem of a shock wave propagating through a non-uniform medium (ideal gas) of decreasing density and reaches a boundary where the density vanishes. For a planar initial density profile of $\rho(x)\propto x^n$, with $x$ the distance from the boundary, a self similar solution exists. \citet{Sapir2011} studied an extension to the problem which includes radiative flux, in the diffusion approximation. At large optical depth from the boundary, the purely hydrodynamical (without diffusion) solution of \citet{Sakurai1960} is applicable, but at $\tau\sim\beta_{\rm sh}^{-1}$, diffusion is significant and must be taken into account (see ~\S\ref{sec:QualitativePicture}). \citet{Sapir2011} present a self similar numerical solution to the problem, assuming constant opacity and radiation dominated gas. This problem, which can describe the planar phase of shock breakout in the absence of wind served as one of several code checks for our program (see \S~\ref{sec:the_code}).

\begin{figure}[h]
\epsscale{1} \plottwo{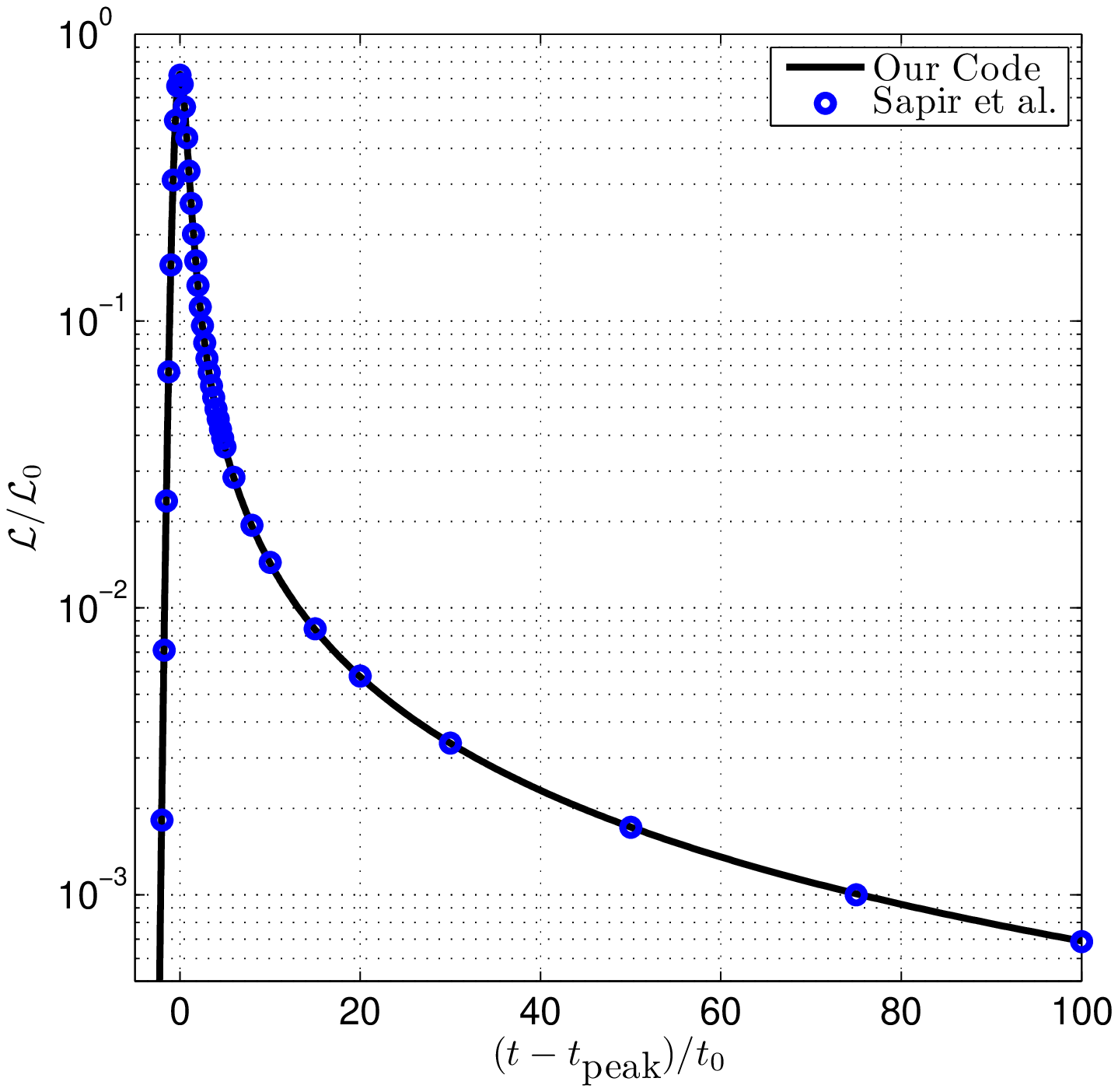}{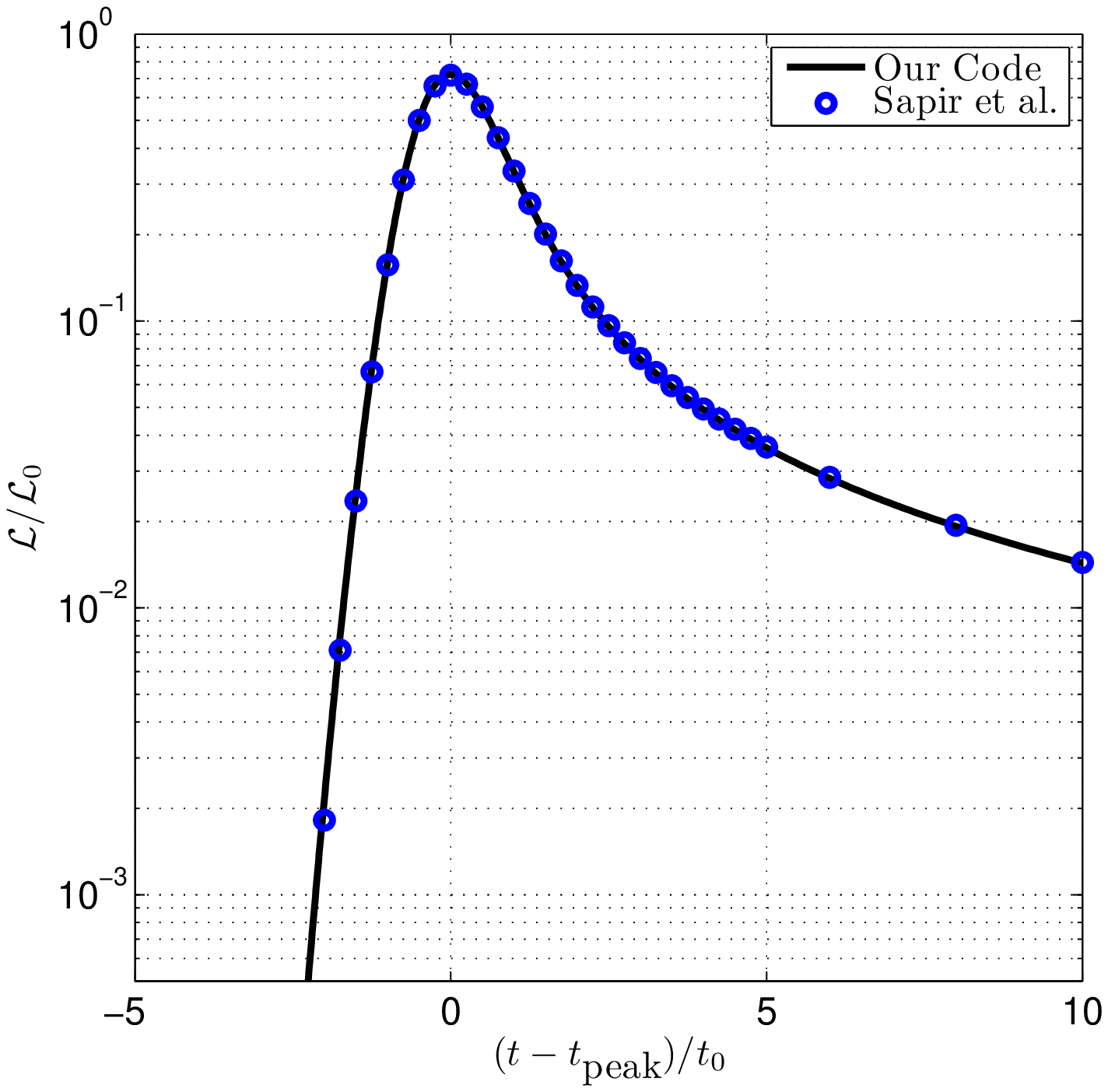}
\caption{Normalized \citep[see][]{Sapir2011} emitted energy flux as a function of normalized time relative to time of peak emitted energy flux, for a density profile with $n=3$. The solid black line is the curve calculated by our code, and the blue circles are taken from table 3 of \citet{Sapir2011}. The right figure is a zoom of the left figure in a shorter time span near the peak. 
\label{fig:sapir}}
\end{figure}

We use planar geometry, insert the appropriate density profile (with $n=3$), and keep only the radiation terms of the EOS (see \S~\ref{sec:the_code}) for the comparison with \citet{Sapir2011}. Our calculation was carried out by moving the inner boundary as a piston at a constant velocity. In addition, we deposited thermal energy in the innermost cell as an initial condition. The shock wave which arises in these conditions converges to the self similar solution \citep{Sakurai1960} as it propagates through the gas. In figure \ref{fig:sapir} we demonstrate the exact fit that we find between our calculated light curve and the self similar solution of \citet{Sapir2011}. Our light curve was normalized to breakout point ($\tau=\beta_{\rm sh}^{-1}$) related values $t_0$ and $\mathcal{L}_0=\rho_0 v_0^3$ \citep[for details see][]{Sapir2011}. 	

\bibliographystyle{apj}

\begin{thebibliography}

\bibitem[Arnett(1996)]{Arnett1996} Arnett, D.\ 1996, Supernovae and Nucleosynthesis, Princeton University Press

\bibitem[Balberg \& Loeb(2011)]{Balberg2011} Balberg S. \& Loeb A.\ 2011, \mnras, 414, 1715

\bibitem[Chatzopoulos et al.(2012)]{Chatzopoulos2012} Chatzopoulos E., Wheeler J.~C \& Vinko J.\ 2012, \apj, 746, 121

\bibitem[Chevalier(1982)]{Chevalier1982} Chevalier, R.~A.\ 1982, \apj, 258, 790

\bibitem[Chevalier \& Irwin(2011)]{ChevalierIrwin2011} Chevalier, R.~A. \& Irwin C.~M.\ 2011, \apjl, 729, L6

\bibitem[Chevalier \& Irwin(2012)]{ChevalierIrwin2012} Chevalier, R.~A. \& Irwin C.~M.\ 2012, \apjl, 747, L17

\bibitem[Elliot(1960)]{Elliot1960} Elliot, L.~A. \ 1960, Proc. R. Soc. Lond. A, 258, 287

\bibitem[Falk \& Arnett(1973)]{FalkArnett1973} Falk, S.~W. \& Arnett W.~D.\ 1973, \apjl, 180, L65

\bibitem[Falk \& Arnett(1977)]{FalkArnett1977} Falk, S.~W. \& Arnett W.~D.\ 1977, \apjs, 33, 515

\bibitem[Gal-Yam(2012)]{GalYam2012} Gal-Yam, A.\ 2012, in preparation

\bibitem[Gezari et al.(2009)]{Gezari2009} Gezari, S., et al.\ 2009, \apj, 690, 1313

\bibitem[Katz et al.(2010)]{Katz2010} Katz, B., Budnik, R. \& Waxman, E.\ 2010, \apj, 716, 781

\bibitem[Katz et al.(2011)]{Katz2011} Katz, B., Sapir, N. \& Waxman, E.\ 2011, preprint {arXiv:1106.1898}

\bibitem[Katz et al.(2012)]{Katz2012} Katz, B., Sapir, N. \& Waxman, E.\ 2012, \apj, 747, 147 

\bibitem[Matzner \& McKee(1999)]{Matzner1999} Matzner, C.~D. \& McKee C.~F.\ 1999, \apj, 510, 379

\bibitem[Miller et al.(2009)]{Miller2009} Miller, A. A., et al.\ 2009, \apj, 690, 1303

\bibitem[Moriya \& Tominaga(2012)]{Moriya2012} Moriya, T. \& Tominaga, N.\ 2012, \apj, 747, 118 

\bibitem[Moriya et al.(2011)]{Moriya2011} Moriya, T., et al.\ 2011, \mnras, 415, 199 

\bibitem[Moriya et al.(2012)]{Moriya2012new} Moriya, T., et al.\ 2012, preprint {arXiv:1204.6109}

\bibitem[Nakar \& Sari(2010)]{Sari2010} Nakar, E. \& Sari, R. \ 2010, \apj, 725, 904

\bibitem[Ofek et al.(2010)]{Ofek2010} Ofek, E.~O., et al.\ 2010, \apj, 724, 1396

\bibitem[Pastorello et al.(2010)]{Pastorello2010} Pastorello, A., et al.\ 2010, \apjl, 724, L16

\bibitem[Quataert \& Shiode(2012)]{QS2012} Quatert, E. \& Shiode J.\ 2012, preprint {arXiv:1202.5036}

\bibitem[Quimby et al.(2007)]{Quimby2007} Quimby, R.~M., et al.\ 2007, \apjl, 668, L99

\bibitem[Quimby et al.(2011)]{Quimby2011} Quimby, R.~M., et al.\ 2011, \nat, 474, 487

\bibitem[Rabinak \& Waxman(2011)]{Rabinak2011} Rabinak, I. \& Waxman E.\ 2011, \apj, 728, 63

\bibitem[Richtmyer \& Morton(1967)]{Richtmyer} Richtmyer, R.~D. \& Morton, K.~W.\ 1967, Difference Methods for Initial-Value Problems, Second Edition, Interscience Publishers

\bibitem[Sakurai(1960)]{Sakurai1960} Sakurai, A.\ 1960, Comm. Pure Appl. Math, 13, 353

\bibitem[Sapir et al.(2011)]{Sapir2011} Sapir, N., Katz, B. \& Waxman, E.\ 2011, \apj, 742, 36

\bibitem[Smith \& McCray(2007)]{SmithMcCray2007} Smith, N. \& McCray, R.\ 2007,\apjl, 671, L17

\bibitem[Smith et al.(2010)]{Smith2010} Smith, N., et al.\ 2010, \apj, 709, 856

\bibitem[Svirski et al.(2012)]{Sari2012} Svirski, G., Nakar, E. \& Sari, R.\ 2012, preprint {arXiv:1202.3437}

\bibitem[von Neumann \& Richtmyer(1950)]{vonNeumann} von Neumann, J. \& Richtmyer, R.~D\ 1950, J. Appl. Phys., 21, 232

\bibitem[Weaver(1976)]{Weaver1976} Weaver, T.~A.,\ 1976, \apjs, 32, 233

\bibitem[Woosley et al.(2007)]{Woosley2007} Woosley, S.~E., Blinnikov, S. \& Heger A.\ 2007, \nat, 450, 390

\bibitem[Zel'dovich \& Raizer(1966)]{Zeldovich} Zel'dovich, Ya.~B. \& Raizer, Yu.~P. 1966, Physics of Shock Waves and High-Temperature Hydrodynamic Phenomena, Dover Publications.

\end{thebibliography}

\end{document}